# Room-temperature polariton repulsion and ultra-strong coupling for a non-trivial topological one-dimensional tunable Fibonacci-conjugated porous-silicon photonic quasi-crystal showing quasi bound-states-in-the-continuum

ATZIN DAVID RUIZ PÉREZ,[1,+] SALVADOR ESCOBAR GUERRERO,[2,+] ROCÍO NAVA,[3] AND JORGE-ALEJANDRO REYES-ESQUEDA[2,4*]

[1] *Universität Paderborn, Warburger Str. 100, 33098 Paderborn, Germany.*
[2] *Instituto de Física, Universidad Nacional Autónoma de México, Circuito de la Investigación Científica, Ciudad Universitaria, Coyoacán, 04510, Ciudad de México, México*
[3] *Instituto de Energías Renovables, Universidad Nacional Autónoma de México, Privada Xochicalco s/n, Temixco, Morelos, 62580, México*
[4] *Sabbatical leave: Département de Physique, Faculté des sciences, Université de Sherbrooke, Québec J1K 2R1, Canada*
[+] *Equal contribution to the paper*
\* *Corresponding author: reyes@fisica.unam.mx*

**Abstract:** Room temperature strong coupling from CdSeS/Zn quantum-dots embedded into a tunable porous-silicon Fibonacci-conjugated array could be observed when exciton's energy was tuned either to the photonic-edge or the defect in the middle of the pseudo-bandgap region of the 1D cavity. Both, the photonic-edge and the defect could be identified as topological edge modes and quasi-bound-states-in-the-continuum, where large density of states and field localization over a wider bandwidth produce a broadband Purcell enhancement, helping to optimize the coupling among the exciton and the 1D photonic quasi-crystal despite the natural difficulty to make the quantum dots to penetrate the cavity pores. A clear repulsion among polaritons, amounting to almost 8 meV for in-plane $k$ values when the cavity energy is larger than the exciton one (blue $k$-detuning), was measured when increasing the incident light fluence, marking the potential of this non-trivial topological array for achieving polariton quantum blockade. Evidence for ultra-strong coupling, where a shift as large as 20 meV was measured, could be found when the defect of the pseudo-bandgap region of the cavity was tuned to the exciton.

## 1. Introduction

The quest for robust quantum technologies, able to outperform classical supercomputers, has driven the recent development of different physical platforms for quantum processors [1,2], being among them superconducting circuits [3], quantum dots [4,5], color centers [6,7], trapped ions [8], and topologically protected systems [9].

On the other hand, ever-present fascination for quasiparticles and nowadays capacity to fabricate, study and apply new solid-state platforms have advanced this quest to the realm of light-matter interaction, where strong coupling of quantum particles to their surroundings, leading to the formation of a quasiparticle called polariton, would allow not only a drastic reduction of future polariton-chips size, but also a drastic increase in their speed performance. Besides, the surge of new physics because of the introduction of non-Hermitian Hamiltonians in polariton systems, is increasing this interest in polaritonics. In particular, exciton-polariton formation at room temperature has become a main target for possible applications due to the possibilities for its manipulation, from polariton lasing to quantum polaritonics and quantum simulations [10-14].

To achieve the exciton-polariton formation, quantum emitters can be coupled to optical microcavities formed by photonic crystals, given their ability for field localization at a given frequency, as well as the control of their dielectric function. However, the very often necessary external support for these microcavities results in a reduction in transmittance and field localization. To compensate for this physical support, a generalized microcavity structure has been previously proposed [15], where an additional Bragg mirror can be conveniently added to the initial microcavity, allowing for a larger transmission, then for a much better photonic response. This use of an added asymmetry to the original symmetric microcavity has suggested that originally asymmetric structures could present also natural advantages over their symmetric counterparts. In that direction, Fibonacci coupled arrays in conjugated symmetry have evidenced perfect light transmission before [15,16]. Actually, some of its generations have an aperiodic structure, while others can be approached to a quasi-mirror symmetry [17]. This symmetry compromise can also be related to the apparition of robust states, with topological nature and called edge modes, where a large density of states and wide field localization are expected [18], giving rise to a broadband large Purcell factor [19]. Indeed, if this edge mode possesses a practically zero bandwidth, then it happens to be also a quasi-bound-state in the radiation continuum (qBIC) [20], favorizing again a large Purcell factor.

Finally, since several years ago, porous silicon (p-Si) has revealed as a very versatile nanoplatform with a coral-like morphology produced by electrochemical etching of crystalline silicon (c-Si) in an aqueous hydrofluoric acid (HF) electrolyte, whose light emission is due to quantum confinement of charge carriers in the nanostructure as p-Si skeleton thickness is reduced to a few nanometers (<4 nm, exciton Bohr radius), with participation of surfaces states [21]. However, its low external quantum efficiency (typically around 10%), wide emission band, and long recombination lifetimes (in the order of milliseconds) make it not so useful for optoelectronic applications. Nevertheless, since the etching process is self-limited and occurs mainly in the pore tips, multilayered porous structures can be produced. The high contrast index between the porous layers, interface quality and large surface area (540-840 $m^2/g$) have been extensively exploited in photonic and biosensing applications [22-26].

In this work, we show how CdSeS/ZnS quantum dots embedded in a 1D Fibonacci-conjugated photonic quasi-crystal microcavity made of porous-silicon, present strong or ultra-strong coupling at room-temperature among the exciton of the quantum dots and the tunable resonance of the photonic array. By choosing the right conjugated-generation of the Fibonacci array, exhibiting the appropriate asymmetry to compensate for the dense Si support and ensure that the defect is placed at the middle of the pseudo-bandgap [15], we selected the adequate fabrication parameters to tune the photonic-edge or the defect of the cavity bandgap to the colloidal quantum dots photoluminescence's peak (achieving a negative cavity-detuning), which showed a shift and a second peak when embedding the quantum dots into the porous structure. A large Rabi splitting of around 68 meV was measured for the strong coupling, climbing until 240 meV for the ultra-strong case, with both being reproduced by using the coupled oscillator model. A systematic decreasing of the emission lifetime was observed at low and high excitation fluences for both cases, as well as emission intensity saturation for high fluences. This reduction of lifetime in a factor of 2 or 3 evidences the large Purcell enhancement provided by the photonic-edge or the defect cavity modes, which are a consequence of the lack of reciprocity of the 1D Fibonacci-conjugated quasicrystal. As well, when increasing the incident light fluence, a clear repulsion among polaritons, amounting to almost 8 meV, for the strong coupling, and to 20 meV, for the ultra-strong case, for in-plane $k$ values when the cavity energy is larger than the exciton one (blue $k$-detuning), was measured, giving insight for a possible polariton quantum blockade.

## 2. Methods and Protocols

*2.1 Porous silicon fabrication*

Porous silicon (p-Si) microcavities were fabricated by electrochemical etching of p-type boron-doped crystalline silicon (c-Si) wafers, with (100) orientation, and electrical resistivity < 0.005 Ω cm. Before the etching process, an aluminum layer was evaporated on the backside of the c-Si wafers and heated up to 550 ºC in an inert atmosphere during 15 min, to make electrical contact. A Teflon cell was filled with an electrolyte composed of aqueous hydrofluoric acid (HF), ethanol and glycerin in a volume ratio of 3:7:1, respectively. The electrochemical etching, where the c-Si substrate was the cathode and a platinum mesh the anode, starts by applying a constant electrical current. The porosity and thickness of a p-Si layer depend on the current density and the etching time, respectively, both controlled by a computer and a Keithley 2450 Source Meter SMU Series. This control of porosity and thickness will determine the optical path length of each layer, allowing then the tuning of its photonic response. To minimize the porosity gradient in each layer, pauses of 1 s every 4 s of etching were implemented during the anodization. P-Si simple bilayers of low and high porosities were then fabricated by alternating the applied current density during the electrochemical etching, among two values, 3 $mA/cm^2$ and 40 $mA/cm^2$, respectively, being denoted as layers A and B. Multiple layer stack samples, or microcavities, were obtained by combining low and high porosity layers in a Fibonacci sequence, as is explained below. After anodization, samples were rinsed with ethanol and dried applying a flow of Nitrogen. Finally, p-Si samples were passivated by thermal oxidation at 300 °C during 30 min. Previously, the layer of aluminum was removed from the c-Si susbstrate to prevent its diffusion during the thermal oxidation, and to avoid possible Al contamination during the quantum dots deposition (see below).

*2.2 Calculation of complex refractive index of porous silicon layers*

P-Si is a nano-structured material composed of a skeleton of c-Si surrounded by air. Its complex refractive index, $\eta = n - ik$, can be calculated by using an effective medium approximation (EMA). Following D. Estrada-Wise and J. A. del Río [27], a Symmetric Bruggeman approximation was used to determine the real part of the p-Si complex refractive index. This EMA considers different sizes of aspherical inclusions embedded in a continuous medium and is applicable to any porosity:

$$P \frac{n_{air}^2 - n_{eff}^2}{n_{air}^2 - 2n_{eff}^2} + (1 - P) \frac{n_{Si}^2 - n_{eff}^2}{n_{Si}^2 - 2n_{eff}^2} = 0 \quad (1)$$

On the other hand, the imaginary part (extinction coefficient) of the p-Si complex refractive index was determined by using the Non-Symmetric Bruggeman approximation:

$$\frac{k^2}{k_{Si}^2} - \frac{k_{air}^2}{k_{Si}^2} = (1 - P)\left(1 - \frac{k_{air}^2}{k_{Si}^2}\right)\left(\frac{k^2}{k_{Si}^2}\right)^{\frac{1}{3}} \quad (2)$$

In particular, the complex refractive index of layers A and B can be calculated using Eqs. (1) and (2), but it is necessary to know the actual porosity (P) of these layers. For this purpose, the gravimetric method [28] was applied on 5 µm thick p-Si films expressly fabricated, where the silicon wafer to be etched is weighted before anodization ($m_1$), immediately after anodization ($m_2$), and after dissolving the pSi layer in an aqueous solution of sodium hydroxide ($m_3$), using the formula:

$$P = \frac{m_1 - m_2}{m_1 - m_3} \quad (3)$$

In this manner, measuring the corresponding samples weight, with a Sartorius Microbalance (model MC 5) with a precision of 0.0005 mg, porosities values of $P_A = 65.1\%$ and $P_B = 86.5\%$ (from standard error propagation, the error percentage is less than 0.3% for each sample porosity), could be estimated for layers A and B, respectively. Additionally, the thickness of the so-formed p-Si films was characterized by using cross-sectional scanning electron microscopy (SEM) images, obtained with a Hitachi S5500 electron microscope. Accordingly, the etch rate of layers A and B are $v_A = 1.72$ nm/s and $v_B = 14.49$ nm/s, respectively. With these values, the time at which each current density needed to be applied to form the desired p-Si thickness of the layers could be determined.

Real and imaginary parts of the complex refractive index of p-Si layers A (black lines) and B (red lines) are shown in Fig. 1. Given its low porosity, layer A has a high refractive index, going from 1.53 at 900 nm, to 1.89 at 400 nm, with an extinction coefficient going from 0.00080 at 900 nm, to 0.112169 at 400 nm. On the contrary, layer B has a low refractive index going from 1.14 at 900 nm, to 1.21 at 400 nm, and an extinction coefficient from 0.00048 at 900 nm, to 0.07236 at 400 nm.

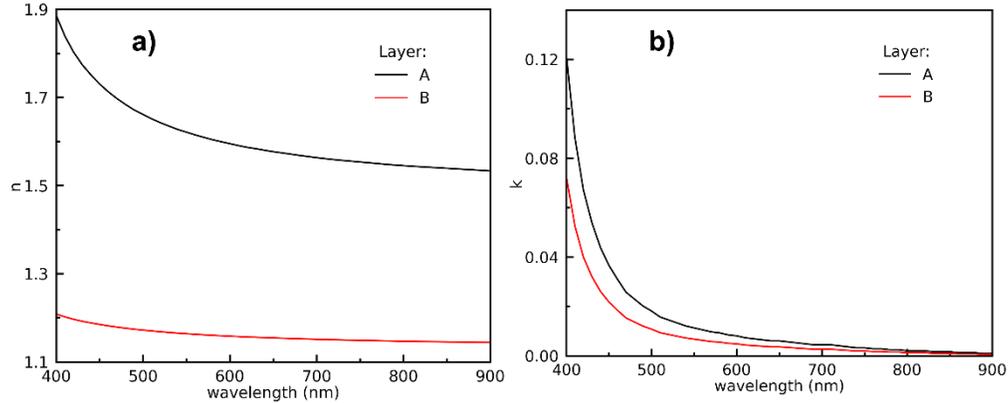

Fig. 1. (a) Real and (b) imaginary parts of the complex refractive index of p-Si layers A (black lines) and B (red lines).

## 2.3 Porous-silicon array in a Fibonacci-conjugated sequences

Porous-silicon array in a Fibonacci sequence is composed of layers A and B, with an optical path length given by

$$n_A d_A = n_B d_B = \frac{\lambda}{4}, \quad (4)$$

where $n$ is the real part of the complex refractive index and $d$ is the thickness of the respective layer. A Fibonacci sequence ($F_j$) is defined following the well-known recursive rule $F_j = \{F_{(j-1)}, F_{(j-2)}\}$, where $j$ is the generation and $j > 1$ [29]. Then, the first Fibonacci generations are $F_0 = \{B\}$ and $F_1 = \{A\}$, such that $F_2 = \{AB\}$, $F_3 = \{ABA\}$, $F_4 = \{ABAAB\}$, and so on. In consequence, the conjugated Fibonacci ($C_j$) is formed exchanging A ⇔ B layers in the same recursive formula, $C_j = \{C_{(j-1)}, C_{(j-2)}\}$, with $C_0 = \{A\}$ and $C_1 = \{B\}$, such that $C_2 = \{BA\}$, $C_3 = \{BAB\}$, $C_4 = \{BABBA\}$, etc.

Therefore, a Fibonacci-conjugated sequence will be formed by joining these so-defined substructures $S_j = \{F_j | C_j\}$, such that one has $S_0 = \{B|A\}$, $S_1 = \{A|B\}$, $S_2 = \{AB|BA\}$, $S_3 = \{ABA|BAB\}$, $S_4 = \{ABAAB|BABBA\}$, etc. The number of layers of the joined array is twice the Fibonacci number (2, 4, 6, 10, 16, . . .). Notice that, in general, there is no internal mirror symmetry with respect to the joining interface, denoted symbolically by (|) [16], this means also that light transmission is not reciprocal, since the cavity is formed by layers A and B with different refractive index and arranged in a quasi-periodic sequence, defining thus that the parity and time reversal symmetries are broken for this structure [18]. This result leads to a non-trivial topological cavity, where, in principle, topological polaritons could be formed [30]. To support this assumption, band structure was calculated by taking all the Fibonacci-conjugated sequence as the unit cell. Reflectance and reflection phase were also calculated and are shown in Fig. 2. Into this figure, for the $n_{th}$ bulk band, it can be observed a discontinuity of the Zak phase, which is a topological invariant associated with 1D photonic systems, and is defined according to the surface bulk correspondence as $\exp(i\theta_n^{Zak}) = -\frac{\text{sgn}(\phi_n)}{\text{sgn}(\phi_{n-1})}$, which can have values of 0 or $\pi$, and $\phi_n$, $\phi_{n-1}$ are the reflection phases in the bandgaps above and below the $n_{th}$ band. This $n_{th}$ band for this quasi-periodic Fibonacci-conjugated sequence corresponds, in this case, to the photonic-edge of the pseudo-bandgap centered around 2 eV, with the photonic-edge located at 1.84 eV (tuned to the exciton of the quantum dots that will be embedded into this cavity and will be shown below), being shown in Figs. 2(a) and 2(b). Fig. 2(c) shows how, for this band, a value of 0 for its Zak phase can be assigned, while for the band above, a $\pi$ value can be identified. This discontinuity in Zak phase introduces an interface state in the gap between these bands, also indicating a band inversion, that is a topological transition [31]. These topological transitions are responsible of the apparition of reflection dips and jumps in the reflection phase within these bands where this discontinuity occurs, as can be also observed in Fig. 2(a) and 2(c). It is worth remarking that these symmetry conditions, that is, the apparition of topological photonic-edge modes, are the result of combining both, Fibonacci, and Fibonacci-conjugated sequences, where this combination is justified in next subsection.

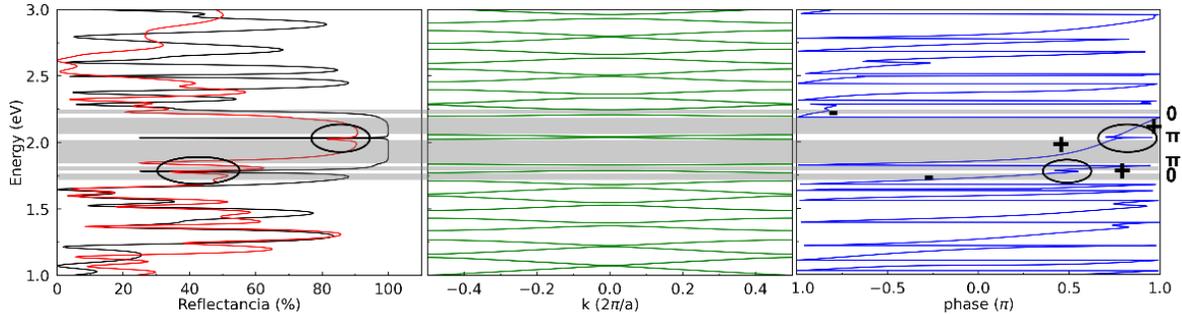

Fig. 2. (a) Reflectance, (b) band structure and (c) reflection phase for the Fibonacci-conjugated quasi-periodic multilayer array. Zak phase discontinuity is illustrated in parts (a) and (c).

*2.4 Reflectance and density of states for asymmetric supported cavities*

As it has been stated in our previous results [15], perfect transmission can be obtained at the bandgap of a photonic system of bilayers when it is composed of a random array of bilayers, given that the number of bilayers AB is equal to the number of bilayers BA, and that there is any number of bilayers AA and BB. As randomness increases, the generation of additional high transmittance states into the array's bandgap results in a reduction of the density of states in the center of the bandgap, with an increasing in its photonic-edge [15], which is an advantage for non-completely homogeneous materials, like p-Si, and asymmetric and aperiodic arrays, like photonic Fibonacci quasi-crystals. This also means a wider field localization at the bandgap photonic-edge, which is an important factor for strong coupling effects, and allows to overcome the negative effects of the rugosity of p-Si materials fabricated by electro-chemical attack for the photonic use of perfect transmission regions [32,33]. When absorption is considered, and the refractive index of the incident media differs from the refractive index of the transmitted media, that is, when the p-Si photonic structure is supported in the Si wafer, the intensity of the localized field may be greatly reduced. This effect can be compensated by adding bilayers to one side of the microcavity, becoming then into an asymmetric microcavity [15], or by using a natural random microcavity, as those obtained for some generations of the

Fibonacci-conjugated sequences. In particular, generation 8, with a layer structure given by $S_8$ = {ABAABABAABAABABAABABAABAABABAABAAB|
BABBABABBABBABABBABABBABBABABBABBA}, fulfills highly random conditions, with a very large and wide density of states (DOS) at the border of its pseudo-bandgap and long-range symmetry for its Fourier spectrum [33]. Fig. 3 shows the calculated and experimental reflectance spectra of $S_8$ array of p-Si layers (p-Si $S_8$), green and black lines, respectively.

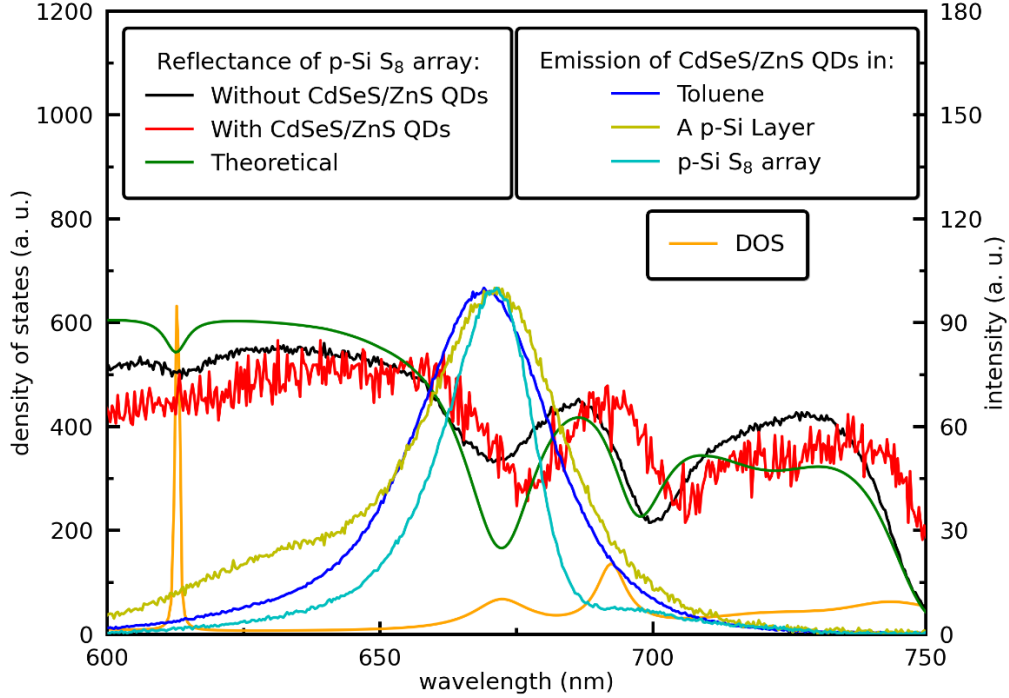

Fig. 3. Reflectance (theoretical and experimental) and DOS for a $S_8$ Fibonacci-conjugated p-Si array. A particularly broad and large DOS can be observed at the photonic band-edge of Fibonacci array, around 672 nm, that is, tuned to CdSeS/ZnS QDs emission in Toluene, which is also shown, as well as QDs emission in a single p-Si layer and the Fibonacci array.

In Fig. 4(a), it is shown a cross-section of $S_8$ p-Si array, obtained with a Hitachi model S-5500 Field Emission Scanning Electron Microscope (FE-SEM). The A layers are identified by being lighter than the B layers. Note that the substructure in Fibonacci sequence ($F_8$) is in the upper part, while the substructure conjugated Fibonacci ($C_8$) is in the lower one. The entire structure is supported on the c-Si substrate, implying strong absorption from the entire system, which makes it naturally non-Hermitian. Fig. 4(b) shows the electric field density for this array, illustrating the field localization mainly at its top surface. Remarkably, as stated above, this specific 1D structure lacks inversion symmetry and reciprocity, in addition to the inherent lack of translational symmetry of the quasicrystals, meaning that parity and time reversal symmetry are broken, which can be translated into possible and robust photonic-edge states, of topological nature [18], as also is discussed in the previous section. The nature of the Fibonacci distribution makes this lack of reciprocity a true time-reversal symmetry breaking. An important aspect to be studied later for this system is if this symmetry breaking, and the resulting topological polaritons, could lead to the long sought backscattering prevention for this system's application in other fields, like quantum technologies.

The large density of states and wide field localization at the photonic-edge of the pseudo-bandgap of this Fibonacci-conjugated array, as shown in Fig. 3, are firm evidence of this fact, which then provides a high Q factor in a wider spectral range (around 80 for the basic definition of Q: $Q = \lambda_{cav}/\gamma_{cav}$, being around 1100 for the defect of the pseudo-bandgap, but larger in both cases when considering the temporal decaying of the electric field into the cavity, and not considering absorption), relaxing the demand on spectral match among this edge mode and the quantum dots' exciton. In consequence, although this topological photonic band-edge mode means a slow light regime, as it has been previously demonstrated for simple Fibonacci quasicrystals [34], it translates into a broadband large Purcell factor [19], as it could be verified by the measurement of lifetime of the formed polariton, which could be orders of magnitude shorter than that of the free exciton [35].

On the other hand, this photonic band-edge mode, with a small bandwidth and such a large field localization for large quasi-crystal, can be also identified as a quasi-bound state in the radiation continuum (qBIC), characterized by a smaller decay rate than free propagation [20,34,36], favorizing again a large Purcell factor. Although counterintuitive, this large factor is the basis for the strong coupling in a system where the QDs are barely physically inside the pores of the Fibonacci quasi-crystal and are far from the region where this field concentration is present, as discussed below for Fig. 5. This identification of the qBIC would mean the formation of polariton BICs or pol-BICs, which are again evidence of the topological charge of this polaritonic system [37,38].

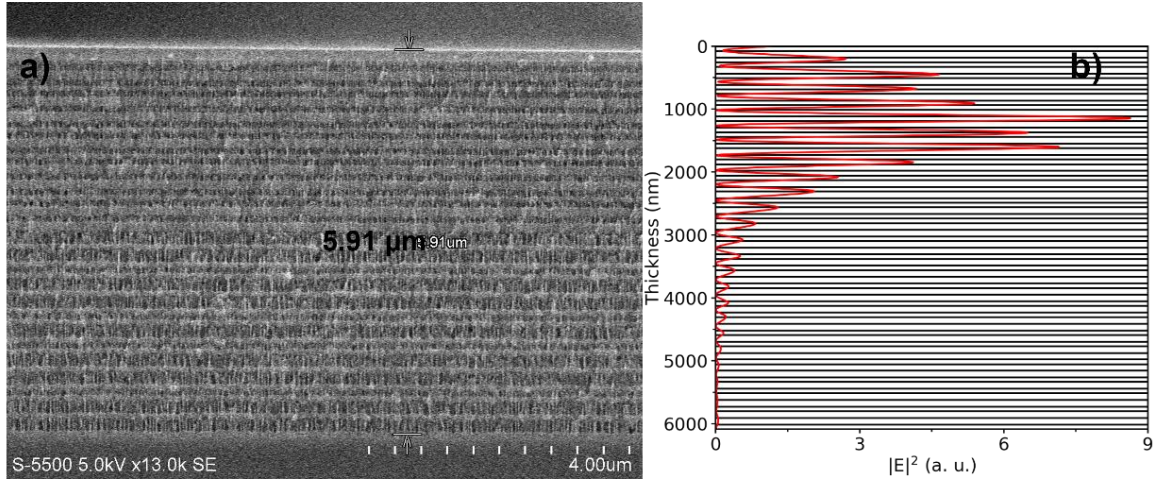

Fig. 4. (a) Cross-section and (b) field localization of $S_8$ Fibonacci-conjugated p-Si array.

*2.5 Quantum dots embedding into the Fibonacci-conjugated array*

Due to their potential for several applications, ranging from biological to quantum computing, semiconductor nanocrystals with a high band gap or quantum dots (QDs) are in trend. Core-shell heterostructures present different localization of charge carriers like electron and holes, due to the spatial alignment occurring in the core and shell levels of the semiconductors when varying the different bandgap energies [39]. Among them, CdSe shows size variation, single-photon emission, and tunable photoluminescence. Its inclusion into a wider bandgap semiconductor shell prevents chemical instability, while charge carriers (electrons and holes) are confined within the narrow band gap semiconductor, thus enhancing the luminescent efficiency resulted by the direct exciton. Also, these QDs have the potential to generate multiple excitons and use them to efficiently convert multiple electron–hole pairs from a single photon, increasing then the possibilities for stronger coupling [39]. All these characteristics make this kind of semiconductor nanocrystals suitable to couple their excitons efficient and strongly to a photonic cavity, as it will be shown below.

Inorganic commercial core-shell Cadmium Selenide Sulfide/Zinc Sulfide (CdSeS/ZnS) QDs from Sigma-Aldrich (753807), with an average size of 6 nm and an emission centered at 672 nm for the specific product bought by our laboratory, have been selected to couple their excitons to the cavity mode of a 1D $S_8$ Fibonacci-conjugated p-Si array. To match the photonic border of the pseudo-bandgap of the array to this emission wavelength, given the measured thicknesses $d_A = 73 \pm 5$ nm and $d_B = 105 \pm 3$ nm, $\lambda = 492$ nm in Eq. 4, in consequence $n_A = 1.67$ and $n_B = 1.17$ (see Fig. 1). The deposition of CdSeS/ZnS QDs into the p-Si $S_8$ array was made by immersing the samples into a QDs colloidal solution. It was prepared by mixing 0.9 ml of toluene with a 0.1 ml of concentrated QDs colloidal solution (concentration of 1.0 nmol/mg) for 24 hours. Then, the samples were taken out from solution and dried in air.

Alternatively, Cadmium Selenide/Zinc Sulfide (CdSe/ZnS) QDs from Cytodiagnostics Inc. (QD-630), with an average size around 10-11 nm and an emission maximum centered at 628 nm for the specific product bought by our laboratory, were also used for achieving strong coupling with the Fibonacci array. All the corresponding parameters and measurements are shown in the Supporting Information file.

Theoretical reflectance was calculated using the classical model of transfer matrix method [40], while the experimental reflectance at normal incidence was measured with a spectrophotometer UV-Vis-IR Shimadzu UV1601, with both being shown in Fig. 3. Furthermore, normalized emission spectrum of CdSeS/ZnS QDs colloidal solution is also shown into this figure with a navy blue line, being excited using a EKSPLA PL2231-50-SH/TH Nd:YAG pulsed laser System, featuring ~26 ps pulses with a repetition rate of 10 Hz. Note the agreement among theoretical and experimental reflectance spectra, and that the maximum of emission of CdSeS/ZnS QDs is well tuned to the border of pseudo-bandgap, that is, to the photonic-edge mode of this 1D topological photonic quasi-crystal. The addition of CdSeS/ZnS QDs in p-Si $S_8$ array gives a modified reflectance spectrum with a red-shift, also shown in this figure, as a result of the strong coupling among the exciton and this cavity mode, and slight increment in the optical path length.

Fig. 5(a) shows the superficial view of p-Si $S_8$ quasi-crystal non-periodic array without CdSeS/Zn QDs, where the p-Si pores can be seen, while Fig. 5(b) shows the array after deposition of the QDs. It can be noticed how QDs have partial or totally covered most of the pores. Figs. 5(c) and 5(d) show the cross-section of a layer of the p-Si $S_8$ array without and with quantum dots, respectively. In both images, a rough (grainy) inner surface of the p-Si can be seen, typical of the material, which makes it difficult to distinguish the quantum dots. This result makes evident that only a small quantity of the quantum dots gets into the pores. Nevertheless, as shown into the results, the QDs are experiencing the large electric field concentration due to the cavity's quasi-BIC given by this topological photonic-edge mode.

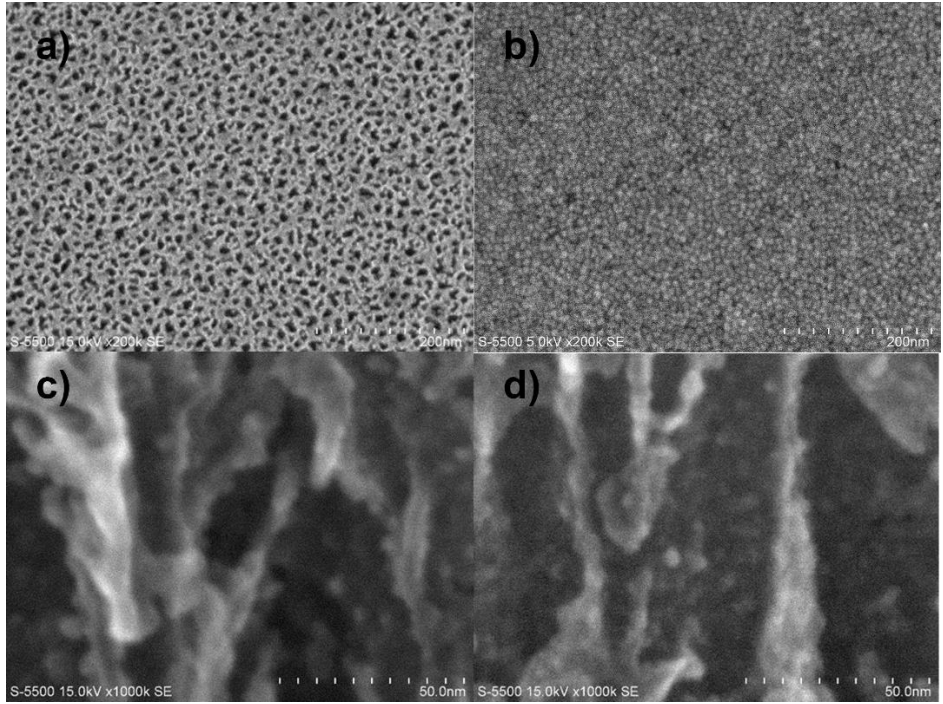

Fig. 5. Superficial view of p-Si S8 array: a) without and b) with CdSeS/ZnS QDs. Cross section of p-Si S8 array: c) without and d) with CdSeS/ZnS QDs

*2.6 Photoluminescence measurement setup*

The angle-resolved emission spectra were measured at the University Laboratory of Optics at Surfaces (Laboratorio Universitario de Óptica de Superficies) into the Physics Institute of UNAM (LOS-UNAM), using the experimental setup shown in Fig. 6(a). The sample was excited at 355 nm by the EKSPLA PL2231-50-SH/TH Nd:YAG pulsed laser System mentioned above. The spot diameter was fixed at 3.6 mm by using a diaphragm. The emission of the sample was collected with an optical fiber (Ocean Optics model P1000-2-UV-VIS with a core of 1000 µm) and analyzed with a UV-VIS spectrometer Ocean Optics USB2000+. Both, the sample and the optical fiber were attached to a rotational plate Newport RSP-1T, to control the incidence and emission angles, respectively.

The emission spectra were obtained by two different processes.
1. The excitation angle was fixed at 60º while the emission angle varied from -32° to 32° in 0.9° steps, i.e., the optical fiber rotates around a fixed sample where the laser beam impinges at 60°. All angles were measured with respect to the samples' normal. This process is represented in Fig. 6(b).
2. The emission angle was fixed at the angle where the anti-crossing point could be observed ($\delta = 0$), which depends on the system. For the sample with the CdSeS/ZnS QDs tuned to the photonic-edge of the Fibonacci array, this angle was 24.3°, as shown below, while the excitation angle varied from 0° to 80°.

To achieve this condition both rotational plates, the one for the optical fiber and the one used for the sample, were rotated at the same time in a clockwise direction. The measurements were done every 0.9 degrees, see Fig. 6(c).

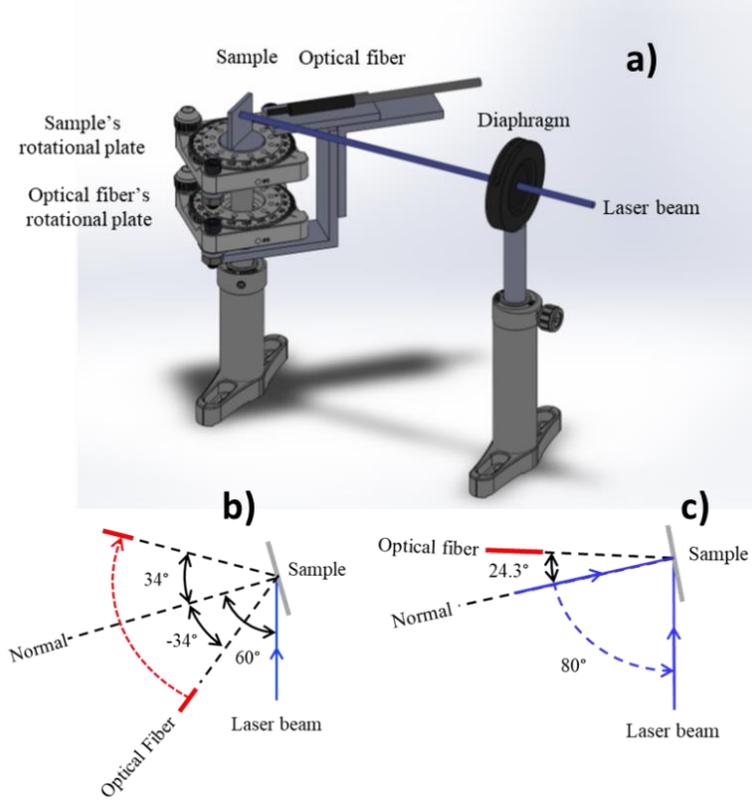

Fig. 6. a) Experimental setup for the emission spectra. The incidence and the recollection angles were measured using rotational plates, while the size of the spot was fixed by using a diaphragm, thus reducing scattered light into the detector, which was coupled to the distal side of the optical fiber. Schematics of the setting up to excite the sample and to collect its emission. b) the excitation angle was fixed to 60°, while the collection angle varied from -34° to 34°, and c) the emission angle was fixed at 24.3°, while the excitation angle varied from 0° to 80°.

## 2.7 Theoretical Analysis

Linear interaction of light with matter can be generally described by a Hamiltonian with one term representing matter corresponding to the electron states for a two-level system, a term corresponding to a structured radiation field and a third term describing the interaction between both systems [41].

In a weak-coupling situation, the interaction term allows to describe photon excitation of an electron in the ground state to the excited one. For a strong-coupling case, this frequent non-Hermitian Hamiltonian can be properly transformed to the framework of the Janeys-Cummings or Tavis-Cummings models, where the structured light field and the specific optical transitions are adequately described by corresponding creation and annihilation operators. Within the context of these quantum pictures, to achieve large coupling strengths, the number of quantum emitters must be increased, or the mode volume confining light has to be decreased [42].

A simpler, more intuitive way of modeling the strong-coupling case is by considering the confined radiation field and the two-level system as coupled oscillators [42,43], which allows then to use the non-Hermitian Hamiltonian given by Eq. (5):

$$H = \begin{pmatrix} \omega_{cav} - i\frac{\gamma_{cav}}{2} & g \\ g & \omega_{ex} - i\frac{\gamma_{ex}}{2} \end{pmatrix}, \quad (5)$$

with the corresponding eigenvalue problem and polaritonic eigenvalues:

$$\begin{pmatrix} \omega_{cav} - i\frac{\gamma_{cav}}{2} & g \\ g & \omega_{ex} - i\frac{\gamma_{ex}}{2} \end{pmatrix} \begin{pmatrix} C \\ X \end{pmatrix} = \omega_{\pm} \begin{pmatrix} C \\ X \end{pmatrix}, \tag{6}$$

$$\omega_{\pm} = \frac{\omega_{cav}+\omega_{ex}}{2} - \frac{i}{2}\left(\frac{\gamma_{cav}}{2} + \frac{\gamma_{ex}}{2}\right) \pm \sqrt{g^2 + \frac{1}{4}\left(\delta - i\left(\frac{\gamma_{cav}}{2} - \frac{\gamma_{ex}}{2}\right)\right)^2}, \tag{7}$$

where $\omega_{cav}$ and $\omega_{ex}$ are the uncoupled Fibonacci cavity mode and QDs exciton angular frequencies, respectively; $\gamma_{cav}$ and $\gamma_{ex}$ are the damping rates of the two states; $\omega_+$ and $\omega_-$ are the frequencies of the hybrid states or, better, the high- and low-polariton branches, respectively. $g$ is the coupling strength, and $\delta = \omega_{cav}(\theta) - \omega_{ex}$ is the angular or in-plane $k$-detuning among the Fibonacci cavity and QDs exciton energies, considering the angle-resolved cavity mode. $C$ and $X$ are the Hopfield coefficients representing the weighting coefficients of the cavity mode and exciton for each hybrid state, where $|C|^2 + |X|^2 = 1$, respectively. The energy separation between the hybrid polariton bands at the anti-crossing, $\delta = 0$, defines the mode splitting or Rabi splitting, given by:

$$\Omega = \sqrt{4g^2 - \left(\frac{\gamma_{cav}}{2} - \frac{\gamma_{ex}}{2}\right)^2}. \tag{8}$$

This non-Hermitian system is not diagonalizable at the critical coupling strength, given by $g_{QEP} = |\gamma_{cav} - \gamma_{ex}|/4$, where the quantum exceptional point of this coupled system will give place to Rabi splitting.

## 3. Results and discussion

As shown in Fig. 3, the emission from the coupled system, 1D p-Si $S_8$ Fibonacci-conjugated array and CdSeS/ZnS QDs, where the emission from CdSeS/ZnS QDs colloidal solution is also shown for comparison, notoriously splits into two peaks, with bandwidths clearly smaller than the peak from the colloidal emission, as a direct result of the QDs embedding into the p-Si Fibonacci-conjugated cavity [44]. This is strong evidence that the coupled system is into the strong coupling regime and, as it will be verified below, the coupling strength is such that there are two spectral peaks. It must be also remarked that this emission from the coupled system, measured at an emission angle of 0° and an excitation angle of 45°, with respect to the normal of the sample, is also well tuned to the border of pseudo-bandgap of the Fibonacci quasi-crystal non-periodic array, as it was for the colloidal solution, assuring then the coupling to the photonic-edge mode of the resonance. Another important aspect to be remarked are the facts that this initial spectrum, obtained in rather standard or normal conditions of photoluminescence measurements, clearly shows the splitting with the more intense peak located towards the smaller wavelength, that is, indicating the position of the high energy polariton branch but, since it is located at larger energies than the photonic-edge mode, also the presence of a negative cavity-detuning ($\Delta = (\omega_{cav} - \omega_{ex})_{\theta=0} < 0$, that is the energy difference among the cavity and the exciton at normal incidence) and, then, the polariton accumulation into the upper branch, as it will be shown shortly in Fig. 9. Transfer-matrix simulations can show that this situation is reversed for positive cavity-detuning.

The normalized emission spectra of p-Si $S_8$ array with CdSeS/ZnS QDs measured at an excitation angle of 60° and at different emission angles are shown in Fig. 7. Unlike plasmon-polariton formation, which can be achieved only for TM incident polarization, in this case, given the 2D isotropy in the plane perpendicular to the 1D Fibonacci quasi-crystal, characteristic of a cavity with parabolic dispersion, strong coupling could be observed for TE and TM polarizations, as well as for right- and left-circular ones, which practically show the same response as the one presented in Fig. 7 for a TE polarized beam. These spectra were obtained at low fluence; however, it will be shown below how polariton-shift changes as a function of fluence, $k$-detuning and incident polarization. Spectra were fitted (dashed green lines) to two Gaussian curves (red and black lines), showing very good agreement with the experimental ones (blue lines). It can also be observed that both peaks shift towards the blue as the emission angle moves from the normal to the sample, see black dashed lines in Fig. 7.

Fig. 8 shows the energy separation of the peaks of the emission spectra as a function of the wavevector (emission angle), for TE, TM, right- and left-circular incident polarizations. There are two local minima in the separation of the peaks, and both occurs when they have approximately the same intensity, 70 meV at -22° and 68 meV at 21°, being not symmetric with respect to 0°; while there is a maximum separation at 0°, 89 meV. This separation increases for angles beyond 28° in both angular directions.

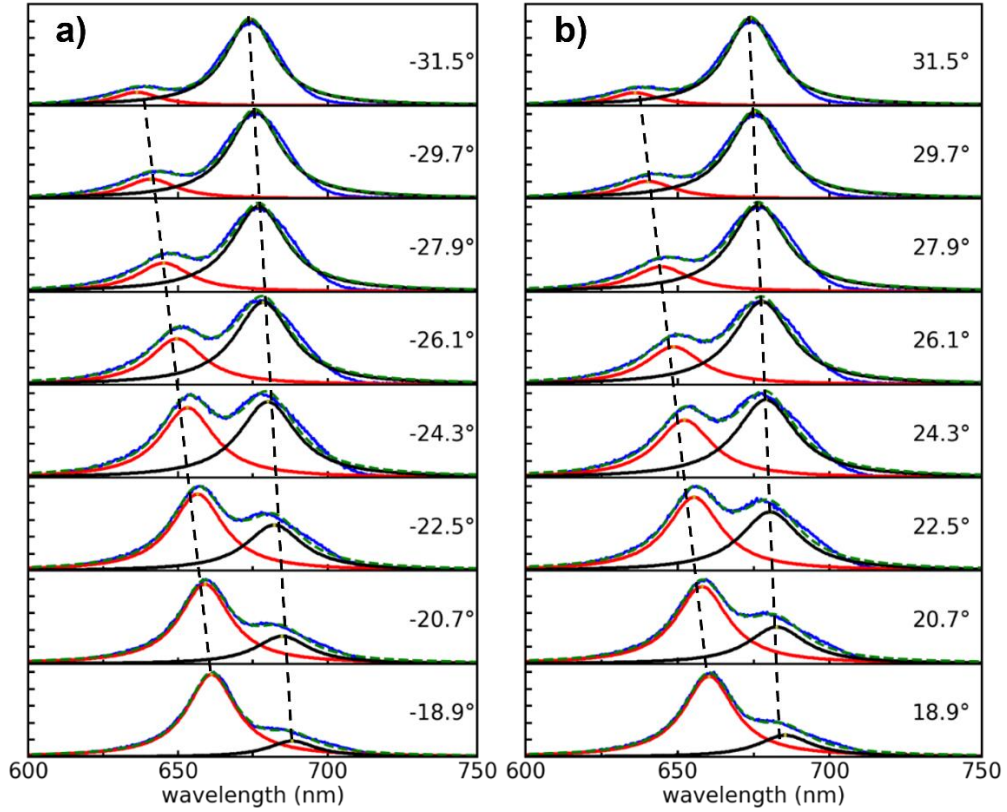

Fig. 7. Normalized emission spectra of p-Si S8 array with CdSeS/ZnS QDs (blue lines) at an excitation angle of 60° and at different and selected emission angles: a) from -18.9° to -31.5° and b) from 18.9° to 31.5°, with 0.9° steps. The excitation beam was TE polarized. The fits (dashed green lines) of two Gaussian curves (black and red lines) are shown. The shift of the peaks of the emission spectra is indicated by black dashed lines.

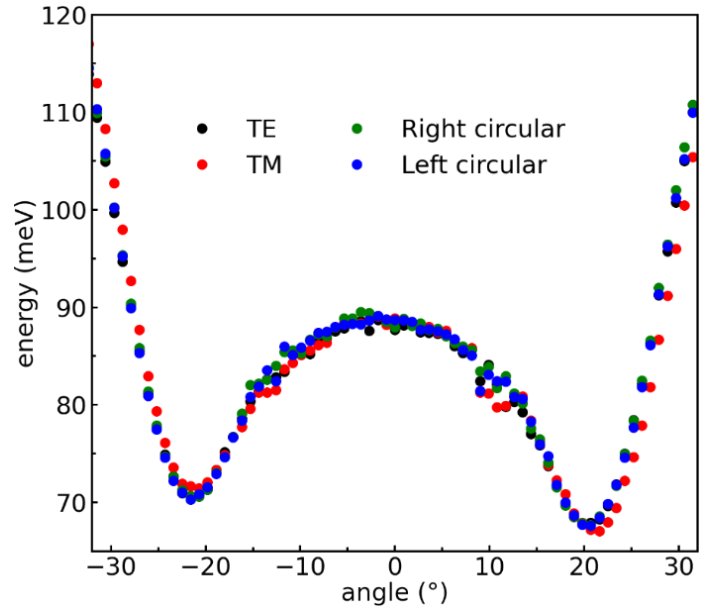

Fig. 8. Energy separation of the peaks of the emission spectra as a function of the emission angle, for the cases of TE (black), TM (red), right- (green) and left-circular (blue) incident polarization.

Extracted PL data from Fig. 8 are shown in Fig. 9, as well as the fitting using the coupled oscillators model presented in Section 2.7, where exciton and cavity dispersions are also shown. From this fitting, at zero $k$-detuning, a Rabi splitting of $67.2 \pm 0.9$ meV could be determined, while measured directly from the data, a value of 67.7 ± 0.2 meV is obtained. Since $\gamma_{cav} = 26.8 \pm 0.4$ meV and $\gamma_{ex} = 72.5 \pm 0.4$ meV, both, $g = 35.5$ meV and $\Omega = 67.2$ meV, satisfy the required conditions for non-vanishing Rabi splitting and spectrally separable resonances, given by $2|g| > \left|\frac{\gamma_{cav}}{2} - \frac{\gamma_{ex}}{2}\right|$ and $\Omega > \frac{\gamma_{cav}}{2} + \frac{\gamma_{ex}}{2}$, respectively. Fig. 10 shows the polariton component from the exciton and photon described by the weighing fractions, given by the Hopfield coefficients as a function of $k$-cavity detuning, $\delta$. The curves are also clearly reflected in Fig. 7, where one can see how the peaks intensities for each polariton, diminish or increase according to the exciton loading given by these coefficients for each in-plane wavevector, indicating the predominance of the exciton emission at each angle, for each polariton branch. A fact that it is not mentioned very often in literature is that the experimental PL intensity decreases at zero $k$-detuning, reaching a minimum, with respect to negative or positive $k$-detuning, when exciton loading of any one of the polaritons increases and becomes maximal, depending on the sign. This fact could be used for detecting the zero $k$-detuning, but it can also be used for other purposes, like sensing.

    An important thing to be observed from Fig. 9 is the clear bottleneck effect present for the low energy branch of the polariton (lower polariton-LP) at this incident fluence [45,46], while the high energy branch (upper polariton-UP) is clearly visible, with the polaritons accumulating at negative cavity-detuning ($\Delta = -76$ meV for the sample whose results are shown in this paper) into this branch. Therefore, for negative cavity-detuning, any polariton interaction will be better observed and characterized in this upper branch, rather than in the lower one, as it will be shown shortly. As the fluence is increased, the bottleneck effect is not overcome, while the UP accumulation is clearer, as it is shown in Fig. S4, for several incident fluences. This result indicates that phonon cooling is not enough to realize a LP condensation at $k=0$. Actually, as it will be shown, LP and UP repulsion prevents this to happen in this system, given that both branches are highly populated, and that both are always present, as shown in Fig. 7.

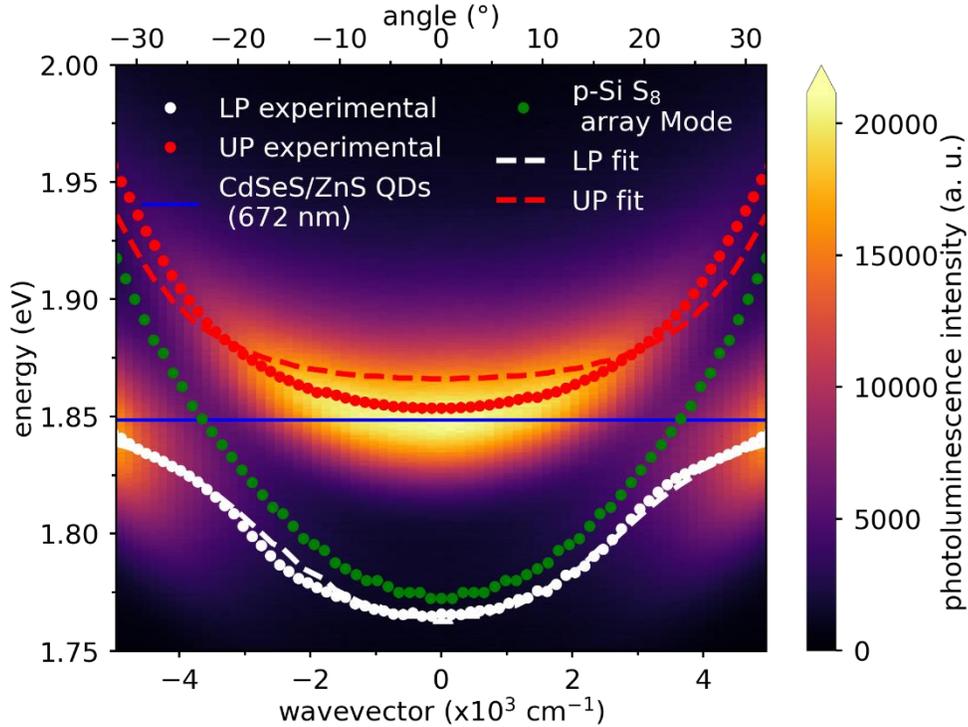

Fig. 9. Formation of exciton-polaritons. The blue solid line indicates the exciton energy of QDs, while the green dots indicate the dispersion of the topological photonic-edge quasi-BIC. Two PL modes are identified as the two anti-crossing polariton branches with dispersion (dashed lines) fitted by a coupled oscillator model with a 67.2 meV Rabi splitting. The red and black dots indicate experimental data extracted from Fig. 8. Polariton bottleneck effect is clearly observed for the LP branch, while polariton accumulation is clearly observed for the upper one, UP.

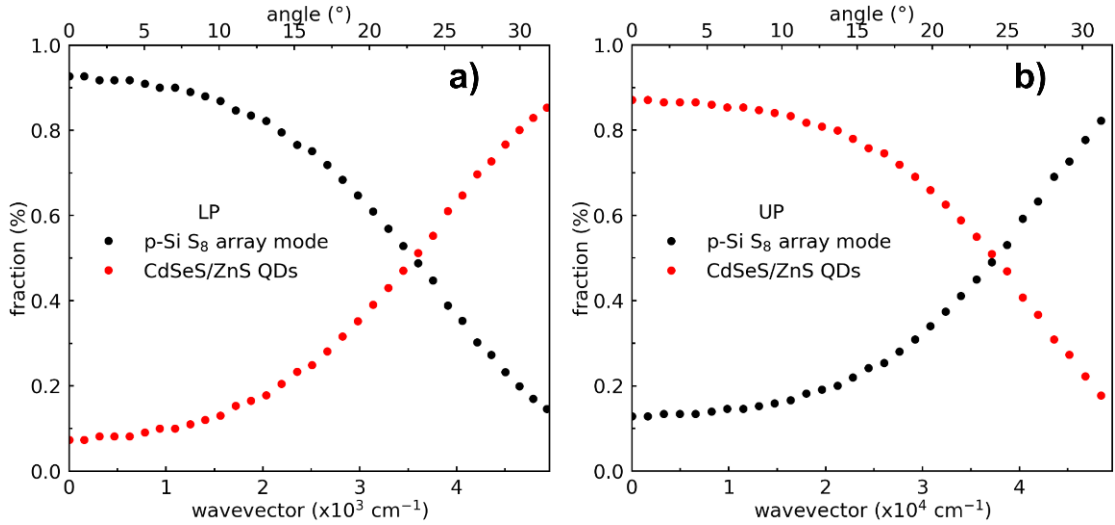

Fig. 10. Hopfield coefficients for the quasicrystal polariton branches, calculated using the coupled oscillator model and providing the weighting of each constituent. The black dots correspond to the coefficients of the topological photonic-edge BIC, while the red dots correspond to those of the QDs exciton.

On the other hand, the normalized angle-resolved emission spectra of p-Si $S_8$ array with CdSeS/ZnS QDs measured by setting the emission angle at negative, zero and positive $k$-detuning, while varying the excitation angle, are shown in Fig. S5, for TE incident polarization. From the fitting of these spectra, it could be observed that UP presents a red shift as the excitation angle increases, while it happens the opposite for LP: it shifts towards the blue with excitation angle, see Fig. 11. The more photonic the UP, the larger this red shift as the excitation angle increases. As already marked for Fig. 7, Fig. 11 also clearly shows that, as the emission angle ($k$-detuning) is increased, upper-polaritons shift to the blue. The most important observation from this figure is that, as the excitation angle is increased, the UP and LP separation decreases, indicating the increasing of their sensibility to the incident light momentum, as its projection into the $k$-plane increases. Fig. S6 shows this for alternative CdSe/ZnS QDs, for unpolarized, TE and TM incident polarizations. From there, it can be also observed that, when the electric field is also projected into the $k$-plane (TM incident polarization), the polaritons get even closer, as if attracted to each other. Although outside the scope of this paper, this result gives insight of possible polariton superfluidity for this coupled system. This topic will be investigated thoroughly soon.

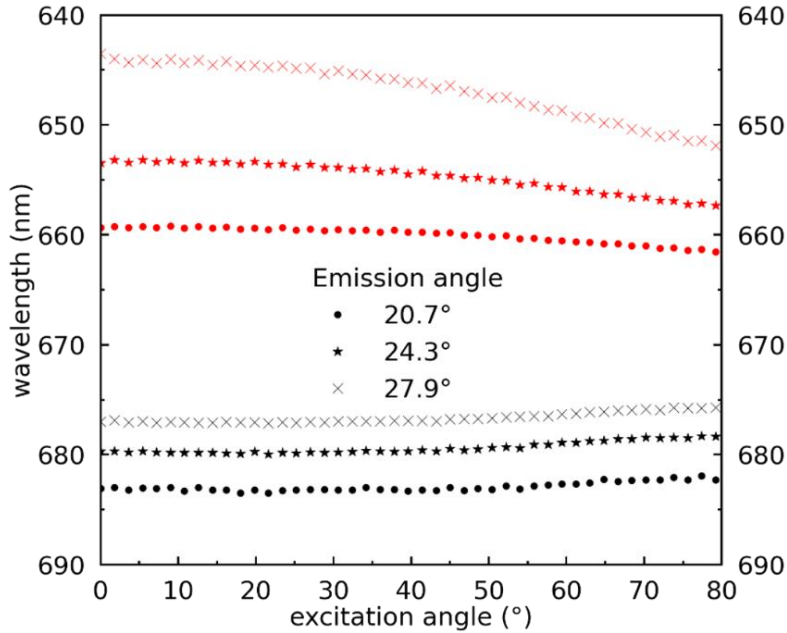

Fig. 11. Polariton position (UP-red, LP-black) for negative (20.7°), zero (24.3°) and positive (27.9°) *k*-detuning when varying the excitation angle.

Now, to study mostly polariton-polariton interaction, that is, polariton nonlinearity, the strong coupling for this cavity-exciton system was examined as a function of the incident fluence. To better estimate this nonlinear interaction, and to compare our results to recent ones, the shift of the polariton energy was measured as a function of the polariton density, which can be estimated from the incident fluence by using Eq. 9:

$$n = (1-R)\frac{E_{total}}{e_p S}, \qquad (9)$$

where the incident energy of the pulsed laser is given by $E_{total} = P/RR$, where $P$ and $RR$ are the incident power and the repetition rate (50 Hz), respectively, with a range of fluences going from 1.13 to 36.30 µJ/cm²; $R$ is given by the experimental reflectance with a value of 0.15, $S$, the spot area, is 6.28 µm², for a laser spot of 0.4 mm in diameter, and $e_p$ is 1.84 and 1.93 meV, for LP and UP, respectively, at low incident fluence. Fig. 12 shows the energy shifting and bandwidth evolution, as well as the intensity saturation, with polariton density, for both polaritons, LP and UP, at TE incident polarization. As mentioned before, given the negative cavity-detuning of the system studied in this work, at normal incidence, the polaritons accumulate on the upper branch rather than in the lower one. Therefore, it is rather normal to expect to observe in this branch what it is normally reported in literature for the lower branch regarding polariton interaction, that is, the blue-shift values. Fig. 12(a) shows LP and UP energy shifting with polariton density for three different *k*-detuning, negative, zero and positive. For the three of them, it can be observed that there is a clear repulsion among LP and UP: LP always shifts towards the red, while UP does towards the blue. This repulsion (|LP shift| + |UP shift|) may amount until 8 meV for the positive case, being the largest, as it has been previously reported in other cases for the blue shifting of the LP [47], indicating the strongest polariton blockade and the largest possible second order correlation measurement for positive detuning. For zero *k*-detuning, both polaritons show a similar slope for their shifting, achieving a maximum repulsion of 6 meV. This corresponds to the similar exciton-photon loading for each polariton, since the system is at its crossing point. Finally, for negative *k*-detuning, the repulsion is only of around 4 meV. A similar behavior was observed for the alternative QDs, CdSe/ZnS, where a repulsion amounting until 9 meV could be observed, see Fig. S8. Regarding the polariton bandwidth, as the polariton density increases, for the UP, in Fig. 12(b), it can be observed the typical broadening related to the exciton-induced dephasing, which is related to the high polariton density for this branch, being the largest for the positive *k*-detuning. On the other hand, for the LP, since the polariton density is lower for this branch in this system with negative cavity-detuning, it can be observed a decrement of the bandwidth as the polariton density increases, being similar for the all the *k*-detuning observed in this part. It is also important to remark that the bandwidth for both polaritons is always smaller than that of the exciton. Finally, Fig. 12(c) allows to corroborate the exciton or photon loading for each polariton as one passes from positive to negative *k*-detuning, despite the increasing of the polariton density, which nevertheless is the reason for the intensity saturation observed in this figure too.

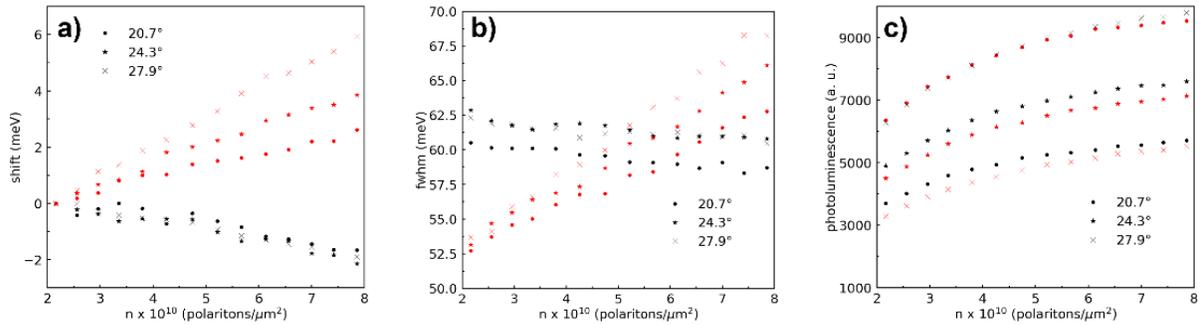

Fig. 12. (a) Energy shifting and (b) bandwidth evolution, as well as the (c) intensity saturation, with polariton density, for LP-black and UP-red polaritons, at TE incident polarization, for negative (20.7°), zero (24.3°) and positive (27.9°) *k*-detuning.

Given the discussion for the observed polariton shifting when varying the excitation angle, where a kind of polariton attraction was observed, being stronger for TM incident polarization, since there is a direct projection of the electric field on the in-plane of the system, the previous results, when increasing the polariton density, were repeated for the same *k*-detuning, positive, zero and negative, but also for TE, TM, right- and left-circular incident polarizations, shown in Fig. 13. For TE polarization, similar results as in the previous paragraph were obtained. For circular polarization, one can roughly say that there is not much difference among right- or left-circular polarizations, being the tendences quite the same for both. However, as observed in [46], the shifting increases drastically more for circular than for linear (TE) polarization. This can be verified in Table 1, where the polariton-polariton interaction constants for each polarization are presented (fitting slopes in Fig. 13 from $\Delta E = gn$, mean-field approximation) and a factor of at least 2 might be identified among linear and circular incident polarizations. From this table, it can be also observed that, for negative and positive $k$-detuning, the most photonic the polariton, the largest the shifting towards the red (LP, negative $k$-detuning) or the blue (UP, positive $k$-detuning). Special attention is deserved for the case of TM incident polarization: except for the UP in positive $k$-detuning, all the shifts are toward the red. This result may be related to the previous one when the excitation angle was varied, given that, for this case, the electric field is projected along the $k$-plane, interacting directly with the exciton dipole. Since Table 2 shows that the slopes are small and quite similar, the $k$-detuning is not a factor limiting this interaction among the electric field and the dipole, and along as the polariton has an exciton loading, the effects of the electric field are similar. Again, this result may be related to possible polariton superfluidity for this coupled system, which will be part of a posterior investigation on this kind of systems.

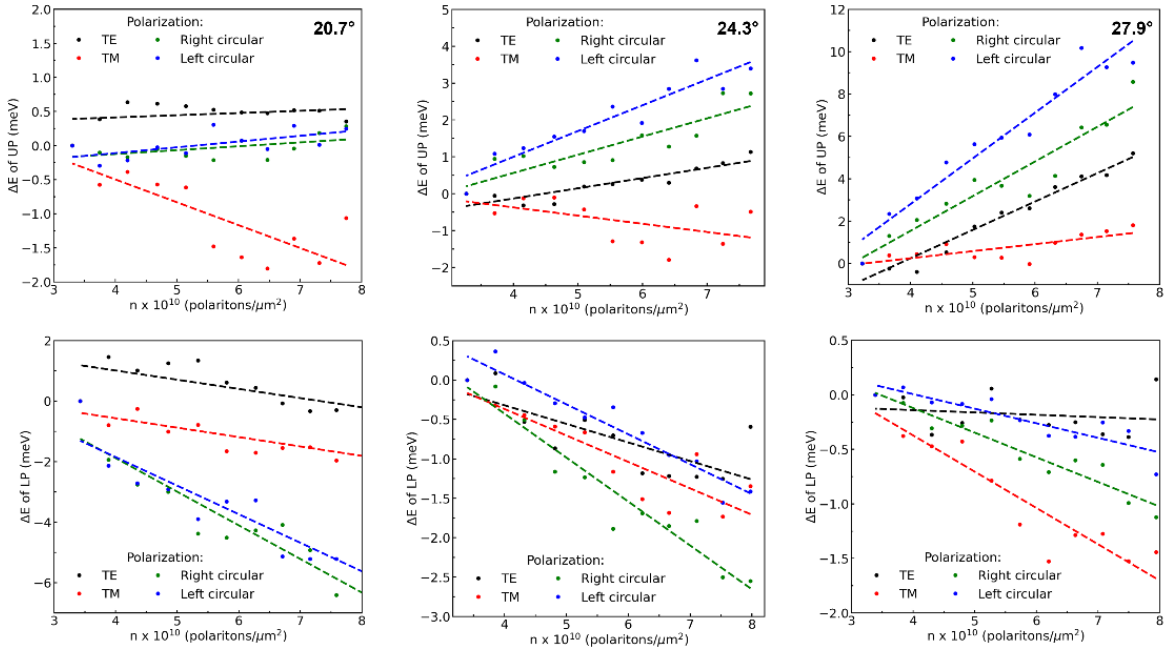

Fig. 13. Energy shifting for LP-black and UP-red polaritons, at TE, TM, right-circular and left-circular incident polarizations, for negative (20.7°), zero (24.3°) and positive (27.9°) $k$-detuning.

Table 1. Polariton interaction constant for different incident polarization for negative (20.7°), zero (24.3°) and positive (27.9°) $k$-detuning.

| $k$-detuning | 20.7 | | 24.3 | | 27.9 | |
|---|---|---|---|---|---|---|
| **Polarization** | **g (feV/µm²)** | | **g (feV/µm²)** | | **g (feV/µm²)** | |
| | **LP (photonic)** | **UP (excitonic)** | **LP** | **UP** | **LP (excitonic)** | **UP (photonic)** |
| **TE** | −30 ± 1 | 3.2 ± 0.1 | −23.6 ± 0.5 | 27.7 ± 0.2 | −2.1 ± 0.2 | 133.9 ± 0.9 |
| **TM** | −31.0 ± 0.8 | −33.5 ± 0.7 | −33.6 ± 0.4 | −22 ± 1 | −33.3 ± 0.2 | 33.3 ± 0.9 |
| **Right-circular** | −111 ± 2 | 5.7 ± 0.1 | −55.8 ± 0.3 | 49.4 ± 0.8 | −22.6 ± 0.1 | 163 ± 3 |
| **Left-circular** | −94 ± 2 | 8.4 ± 0.1 | −38.0 ± 0.2 | 70.0 ± 0.7 | −13.4 ± 0.1 | 215 ± 3 |

Rabi splitting also increases linearly with polariton density, as shown in Fig. 14, for TE and circular incident polarizations, where it can be observed that Ω values are larger for circular than for linear (TE) polarization. Mention apart is done again for TM polarization, since larger and rather independent values could be observed for the polariton densities studied in this work.

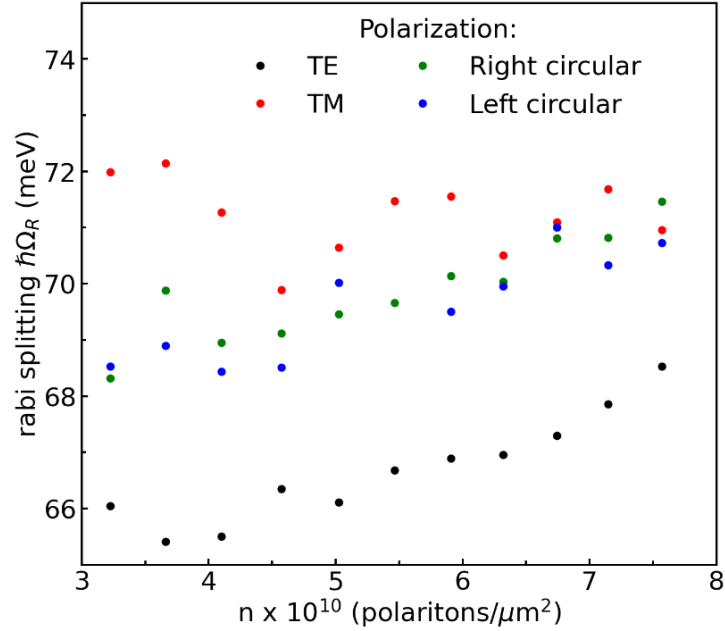

Fig. 14. Rabi splitting as a function of polariton density at TE, TM, right-circular and left-circular incident polarizations.

Polariton lifetime measurements were also performed for these three different $k$-detuning, negative, zero and positive. First of all, for this system, CdSeS/ZnS, and for the alternative one, CdSe/ZnS, is evident the decreasing of the lifetime for both polaritons, compared to that of the QDs' excitons, due to the strong coupling of the excitons to the Fibonacci photonic-edge mode, which corresponds to a quasi-bound-state-in-the-continuum, being characterized by a large Purcell effect and high-quality factor. These results make evident the flexibility of this topologically non-trivial Fibonacci cavity, since it can be easily tuned when fabricated to couple to any kind of emission system, supports high fluences without apparent degradation and allows strong coupling, at room-temperature, even when most of the emitters may be not embedded at their pores. On the other hand, when measured for increased fluences, despite the measurement incertitude, in general, lifetime decreased with fluence, which is in concordance with the tendence showed for Rabi splitting in previous figure. All these results are quite better illustrated for the alternative QDs, CdSe/ZnS, as shown in Fig. S7.

Finally, a new Fibonacci array was fabricated, but this time, the defect in the middle of the pseudo-bandgap was tuned to the exciton resonance, as shown in Fig. 15. Again, a splitting of the photoluminescence can be observed, but this time, the new peak appears at a higher energy, towards the blue. As the emission angle is varied, one can observe that there is a smaller energy exchange among the polaritons (Fig. 16(a)), and that a zero $k$-detuning, where both polaritons have the same intensity or loading, is not evident. It looks like LP is always excitonic in nature, with UP being rather photonic all the time. Then, the dispersion relation gives insight of these facts by showing (Fig. 16(b)) that LP has a negative effective mass and, when calculating the Rabi splitting by using the coupled oscillators model, Ω has a value of around 240 ± 2 meV, which is around 13% of the exciton bandgap. The negative effective mass is evidence of a dissipative coupling, which leads to level attraction [48-50]. This clear negative mass for this effect can also be slightly evidenced in Fig. 9 for the LP at angles around the crossing point, where an inflexion point of the curvature of the LP can be observed. This result combined with those from Fig. 10, which were similar for this new Fibonacci array, where polariton attraction is evidenced, shows the presence of dissipative coupling for this system, which combined with coherent coupling gives a larger control of the effective light-matter interaction, and, as also mentioned in [48], allows the study of polariton superfluidity with polaritons going in the direction or in the opposite one to an applied force. The large value of Ω shows the presence of ultrastrong coupling when this tuning of the defect of the bandgap and exciton emission is achieved. This is a consequence of the smaller bandwidth of this defect in the bandgap (around 20 mev) and then a larger quality factor. In fact, as for the photonic-edge of the bandgap, this band is also a quasi-bound state in the continuum and, since Zak phase presents the same behavior as for the photonic edge, it is also a non-trivial topological band, but with better photonic characteristics than the photonic-edge, being this the reason for observing the ultrastrong coupling. However, since the bandwidth is smaller, this narrowness of the defect makes more complicated the tuning with the exciton resonance.

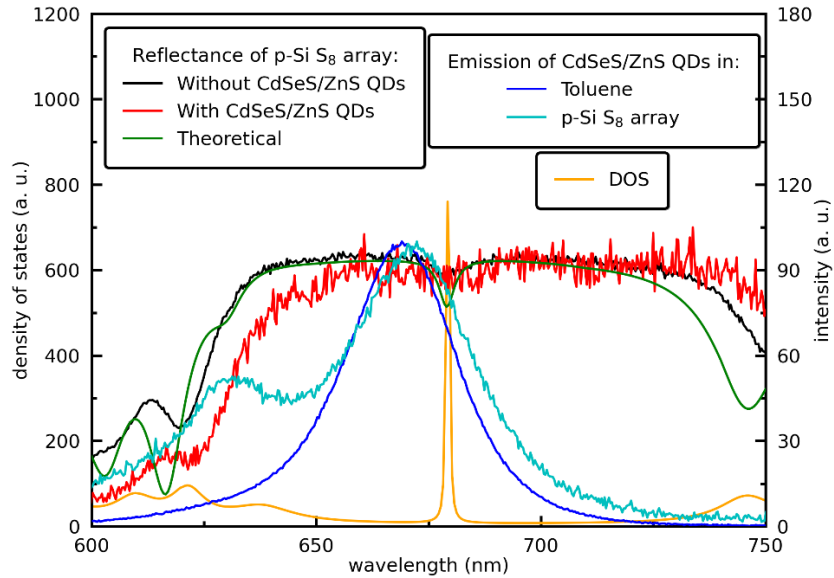

Fig. 15. Reflectance (theoretical and experimental, with and without CdSeS/ZnS QDs) for a $S_8$ Fibonacci-conjugated p-Si array. A very narrow and large DOS can be observed at the defect in the pseudo-bandgap of Fibonacci array, around 672 nm, that is, tuned to CdSeS/ZnS QDs emission in Toluene, which is also shown, as well as QDs emission in the Fibonacci array, showing the emission splitting due to the strong coupling among the exciton and this bandgap's defect.

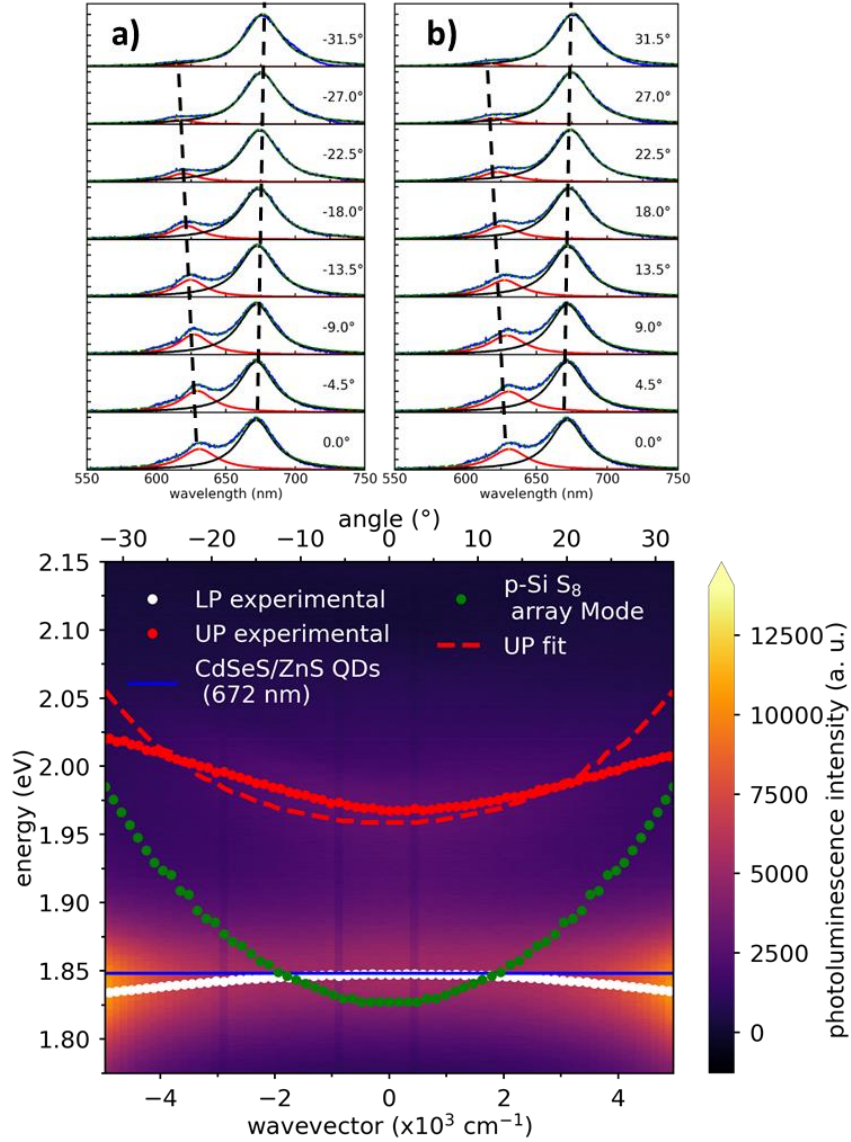

Fig. 16. (a) Normalized emission spectra of p-Si S8 array with CdSeS/ZnS QDs (blue lines) at an excitation angle of 60° and at different and selected emission angles: a) from -18.9° to -31.5° and b) from 18.9° to 31.5°, with 0.9° steps. The excitation beam was TE polarized. The fits (dashed green lines) of two Gaussian curves (black and red lines) are shown. The shift of the peaks of the emission spectra is indicated by black dashed lines. (b) Formation of exciton-polaritons. The blue solid line indicates the exciton energy of QDs, while the green dots indicate the dispersion of the topological defect quasi-BIC. Two PL modes are identified as the two anti-crossing polariton branches with dispersion (dashed lines) fitted by a coupled oscillator model with a 240 meV Rabi splitting. The red and black dots indicate experimental data extracted from part (a).

Regarding the polariton-polariton interaction, first, as for the photonic-edge, the Rabi splitting increases with fluence, Fig. 17(a), but this time it starts to show saturation for all the incident polarizations. For the energy shifting and bandwidth evolution, showed in Figs. 17(b) and 17(c), it is evident the polariton-polariton repulsion, which amounts until more than 20 meV. It must be noted that, this time, LP shifting to the red is almost negligible, but its negative dispersion, coupled with repulsive polariton–polariton interactions, leads to the possible formation of bright solitons [51].

Finally, for lifetime measurements, as for the photonic-edge, it decreases for both polaritons, compared to that of the QD's excitons, and when increasing fluence, follows the tendence for the growing of Rabi splitting with fluence.

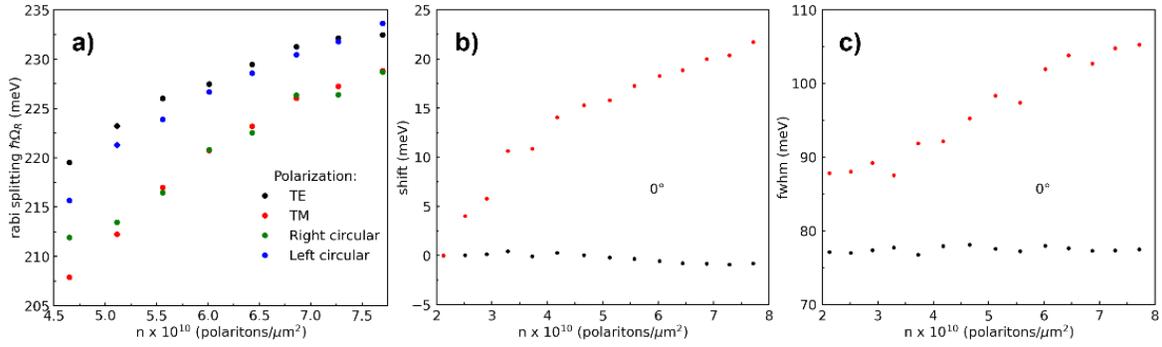

Fig. 17. (a) Rabi splitting as a function of polariton density at TE, TM, right-circular and left-circular incident polarizations. (b) Energy shifting and (c) bandwidth evolution with polariton density, for LP-black and UP-red polaritons, at TE incident polarization, for 0°, that is, negative $k$-detuning.

## 4. Conclusions

From the results of this work, several things can be concluded. First at all, it is necessary to remark the simplicity of the porous silicon photonic structure used to obtain these results. The structure has been fabricated without any special characteristic conditioning the strong coupling, at room temperature, of the photonic cavity with the two different sets of quantum dots studied in this work. In that sense, there is still room for improving the porosity of these photonic structures, then their interaction with the excitonic systems and the consequent strong coupling among them.

On the other hand, two different approaches show the topological nature of this Fibonacci-conjugated quasi-crystal, the Zak phase discontinuity, and the formation of pol-BICs due to the identification of the photonic-edge (or the defect in the middle) of the bandgap as quasi-bound-states-in-the-continuum. The consequent large Purcell effect and high-quality factor would be behind of the strong and ultra-strong coupling observed for this system, as well as for the reduction of its lifetimes when strongly coupled, even if the QDs are rather in the surface of the porous cavity than into it. It is still necessary to determine if this non-reciprocal topological nature for this system is enough to assure no backscattering from it, to then be used in integrated photonic devices. Even more importantly, it would be the use of these symmetries to create Majorana polaritons. From this part, it has to be remarked how the lack of strong order, a characteristic of this aperiodic photonic structure, is used to reinforce topologically the strong coupling, that is, the photonic response of the coupled system.

However, the most important result from this work is the polariton-polariton interaction, where a very large repulsion has been shown for both, the strong and the ultrastrong coupling regimes, as a function of the incident fluence, being the largest for a positive $k$-detuning. This repulsion could be modulated by varying the excitation angle and controlling the incident polarization, where a TM excitation has shown interesting polariton attraction. Both conditions, as the high-Q factors, could be exploited to form controlled superfluid polariton currents or obtaining a polariton quantum blockade. Therefore, the proper control of exciton and cavity energies allow tailoring polariton lifetimes and effective mass, allowing the formation of integrated non-Hermitian and topological polaritonic devices working efficiently at room temperature. In this direction, the use of emitters with a characteristic spin loading, as inorganic perovskites and NV-centers in nanodiamonds, with the tuning facility of porous silicon photonic structures, as well as their symmetric properties, is a work in progress to show the control of the cavity-detuning as a preparation for a larger effort to obtain polariton blockade and superfluidity, which would make these systems a strong proposition for topological quantum technologies.

**Acknowledgements.**

This research was partially funded by ECOS-Nord CONACyT-ANUIES 315658, PAPIIT-UNAM IN112022 and PAPIIT-UNAM IN109122. J.A.R.E. thanks sabbatical funding from PASPA-UNAM, CONACyT and University of Sherbrooke. A.D.R.P. thanks SECTEI CdMx for posdoctoral fellowship SECTEI/120/2022. S.E.G. thanks CONACyT for postdoctoral fellowship. The authors wish to acknowledge the technical assistance of Gerardo Daniel Rayo López and José Campos Álvarez for SEM images.

# Supporting Information

# Room-temperature polariton repulsion and ultra-strong coupling for a non-trivial topological one-dimensional tunable Fibonacci-conjugated porous-Silicon photonic quasi-crystal showing quasi bound-states-in-the-continuum

ATZIN DAVID RUIZ PÉREZ,[1,+] SALVADOR ESCOBAR GUERRERO,[1,+] ROCÍO NAVA,[2] AND JORGE-ALEJANDRO REYES-ESQUEDA[1,3*]

[1] Instituto de Física, Universidad Nacional Autónoma de México, Circuito de la Investigación Científica, Ciudad Universitaria, Coyoacán, 04510, Ciudad de México, México
[2] Instituto de Energías Renovables, Universidad Nacional Autónoma de México, Privada Xochicalco s/n, Temixco, Morelos, 62580, México
[3] Sabbatical leave: Département de Physique, Faculté des sciences, Université de Sherbrooke, Québec J1K 2R1, Canada
[+] Equal contribution to the paper
* Corresponding author: reyes@fisica.unam.mx

**Supporting Note 1:** Alternative Cadmium Selenide/Zinc Sulfide (CdSe/ZnS) QDs

To match the border of the pseudo-bandgap of the array to this emission wavelength, given the measured thicknesses $d_A$ = 60 ± 5 nm and $d_B$ = 95 ± 3 nm, $\lambda$ = 464 nm in Eq. 4, in consequence $n_A$ = 1.93 and $n_B$ = 1.23. In Fig. S1, it is shown the calculated and experimental reflectance spectra of p-Si $S_8$ array, with blue and black lines, respectively.

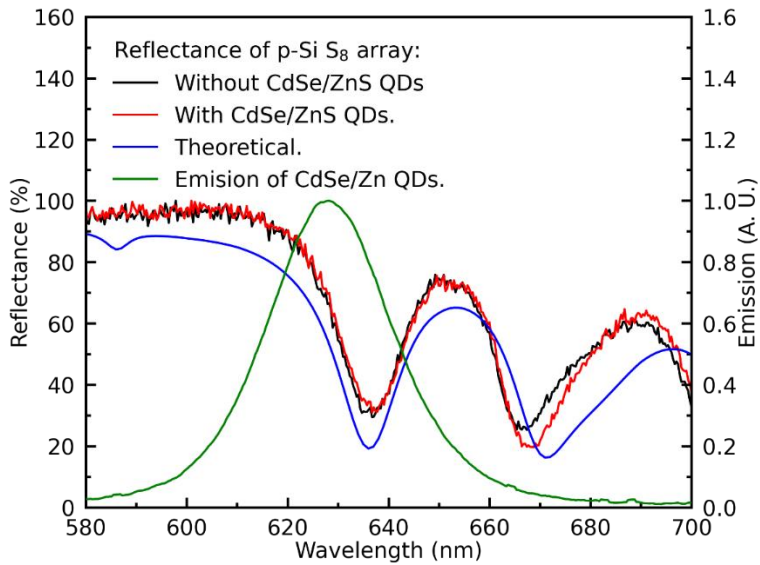

Fig. S1. Theoretical (blue line) and experimental reflectance spectra of p-Si $S_8$ array without (black line) and with (red line) CdSe/ZnS QDs. Furthermore, emission spectrum of CdSe/ZnS QDs colloidal solution is shown (green line).

Fig. S2 compares the emission from the coupled system, when using the CdSe/ZnS QDs, to the colloidal emission

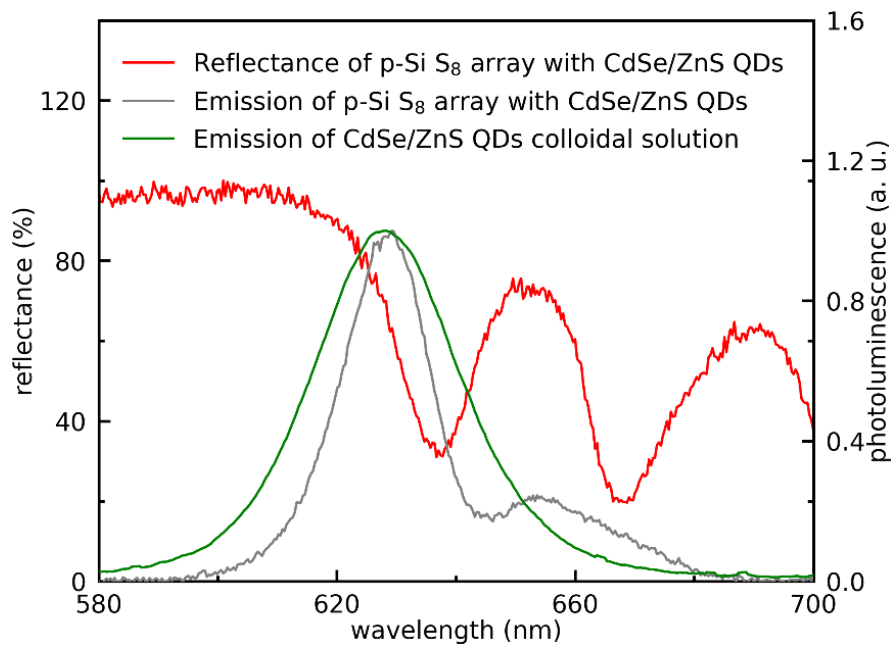

Fig. S2. Normalized emission spectra of Fibonacci array doped with CdSe/ZnS QDs, measured at an emission angle of 0º and an excitation angle of 45º, and of CdSe/ZnS QDs colloidal solution. The reflectance of the coupled system, measured at 0º approximately, is also shown.

Fig. S3 shows emission spectra at different emission angles for CdSe/ZnS QDs.

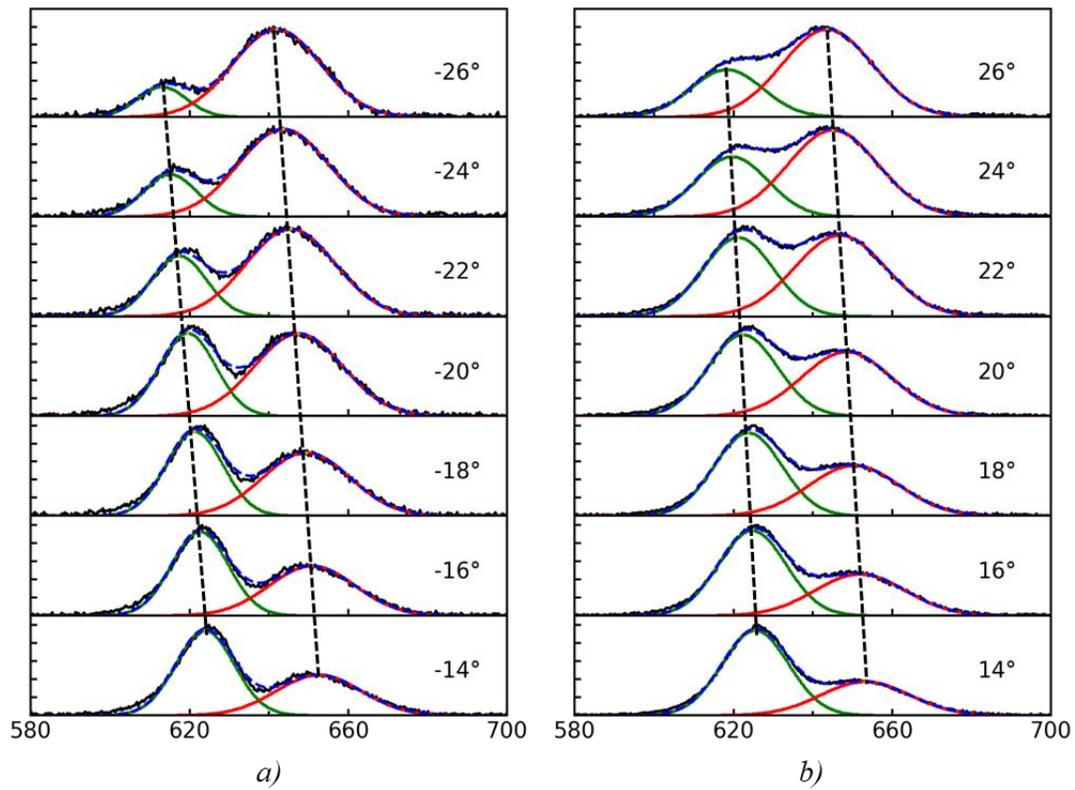

Fig. S3. Normalized emission spectra of p-Si S8 array with CdSe/ZnS QDs (black lines) at an excitation angle of 60° and at different and selected emission angles: a) from -14° to -26° and b) from 14° to 26°, with 2° steps. The excitation beam was not polarized. The fits (dashed blue lines) of two Gaussian curves (green and red lines) are shown. The shift of the peaks of the emission spectra is indicated by black dashed lines.

Fig. S4 shows strong coupling fitting as the fluence is increased, and when TE or TM incident polarizations are used.

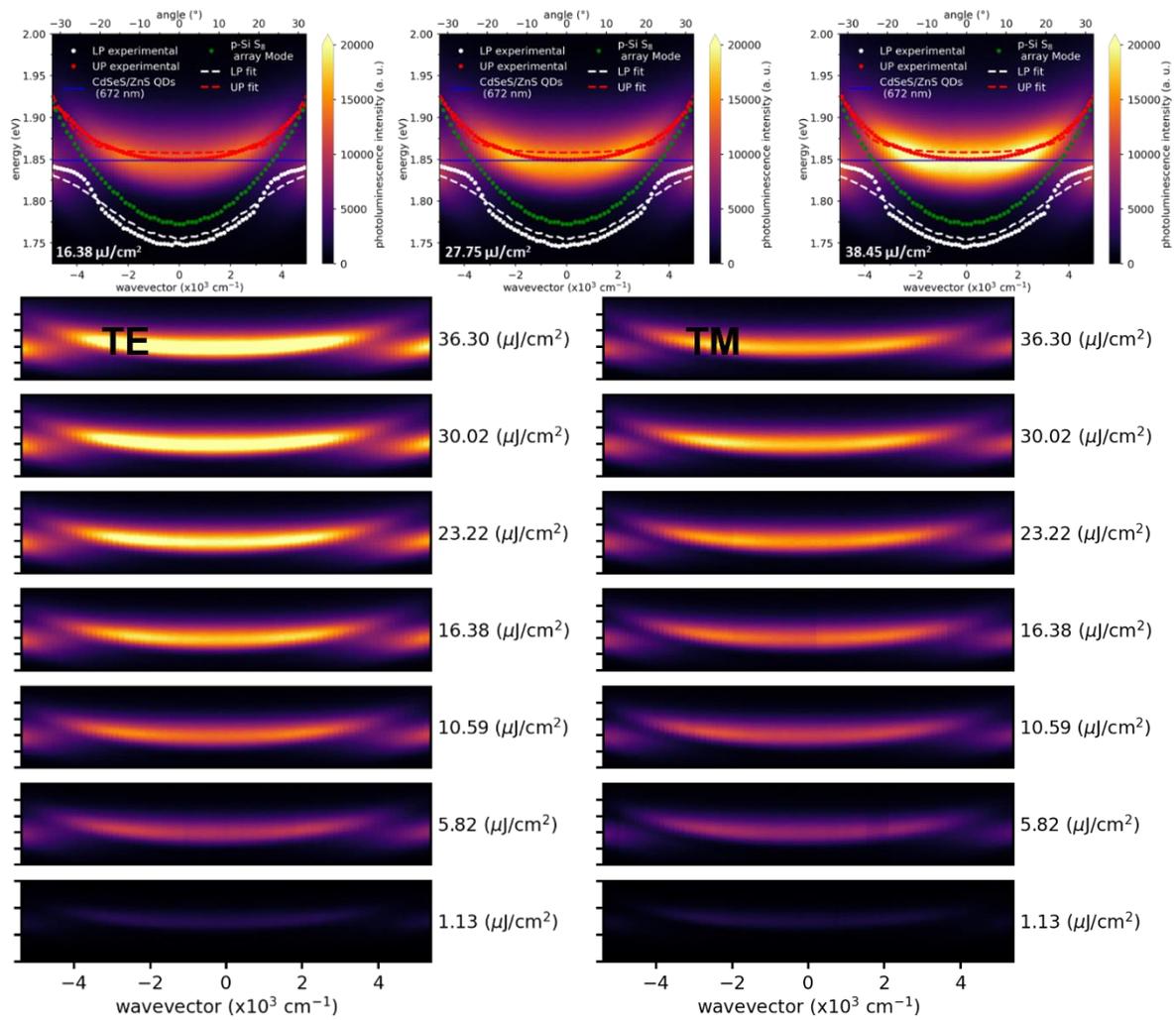

Fig. S4. Strong coupling at three different fluences and for TE and TM incident polarizations.

Fig. S5 shows emission spectra for CdSeS/ZnS QDs at three different emission angles when varying the excitation angle.

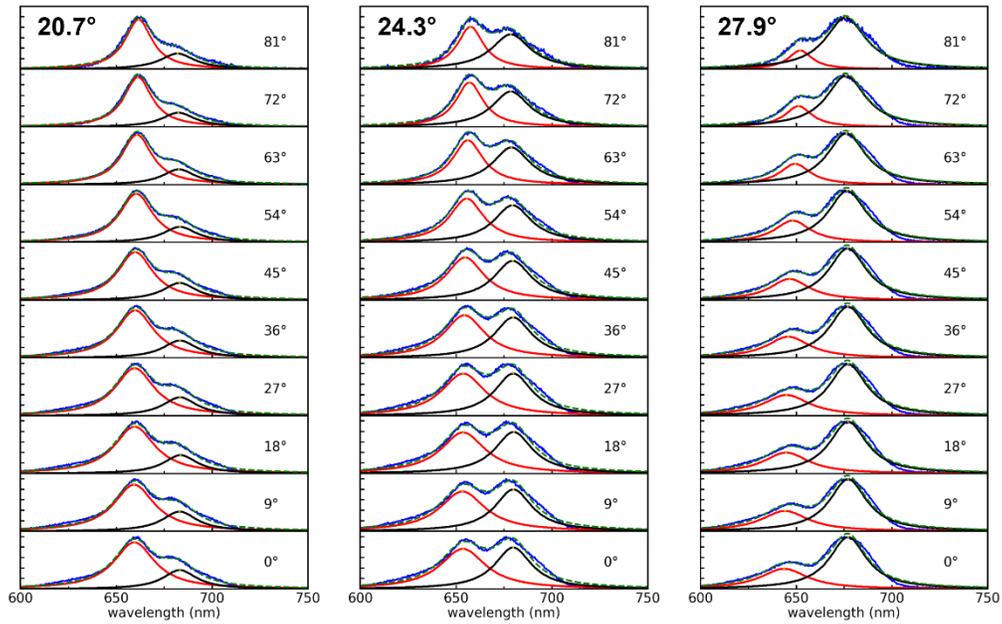

Fig. S5. Normalized emission spectra of p-Si S8 array with CdSeS/ZnS QDs at three emission angles representing negative, zero and positive *k*-detuning, while varying the excitation angle. The excitation beam was TE polarized.

Fig. S6 shows polariton separation for CdSe/ZnS QDs when varying the excitation angle, for different incident polarizations.

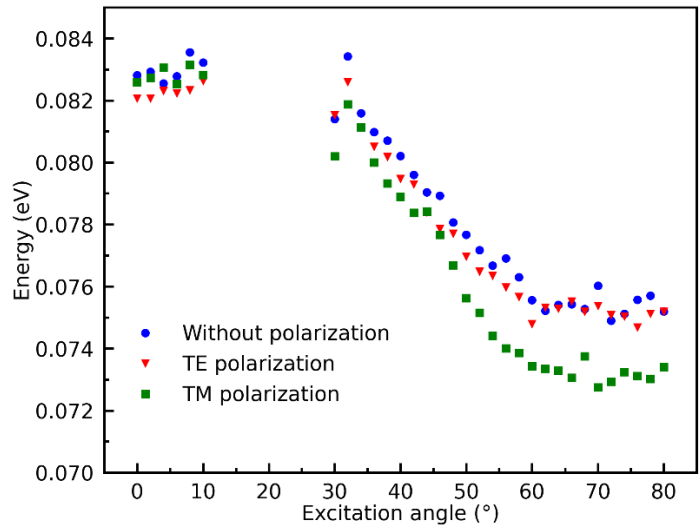

Fig. S6. Polariton separation for p-Si S8 array with alternative CdSe/ZnS QDs, for different incident polarizations. It is worth mentioning that emission could not be measured at excitation angles between 12° and 28° since the laser reflection saturated the detector.

Fig. S7 shows lifetime measurement for CdSe/ZnS QDs at zero *k*-detuning.

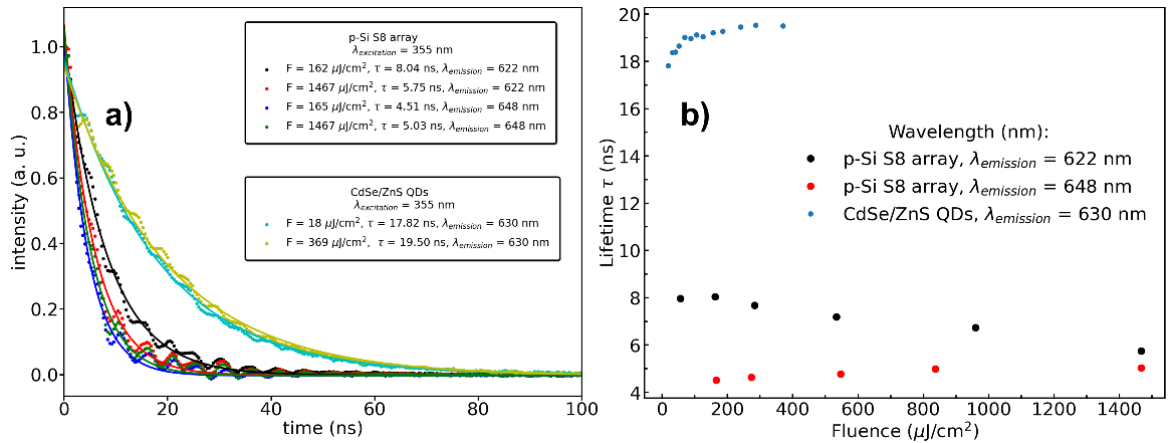

Fig. S7. a) Lifetime measurement of QDs and polariton systems, using non-polarized, ps-pulsed excitation at 355 nm, at peak wavelengths of colloid QDs, lower- and upper-polaritons emission, respectively, for low and high fluences, at zero $k$-detuning. b) Lifetime as a function of fluence for colloid QDs and both polariton branches.

Fig. S8 shows emission spectra at negative $k$-detuning and polariton repulsion for CdSe/ZnS QDs as a function of fluence.

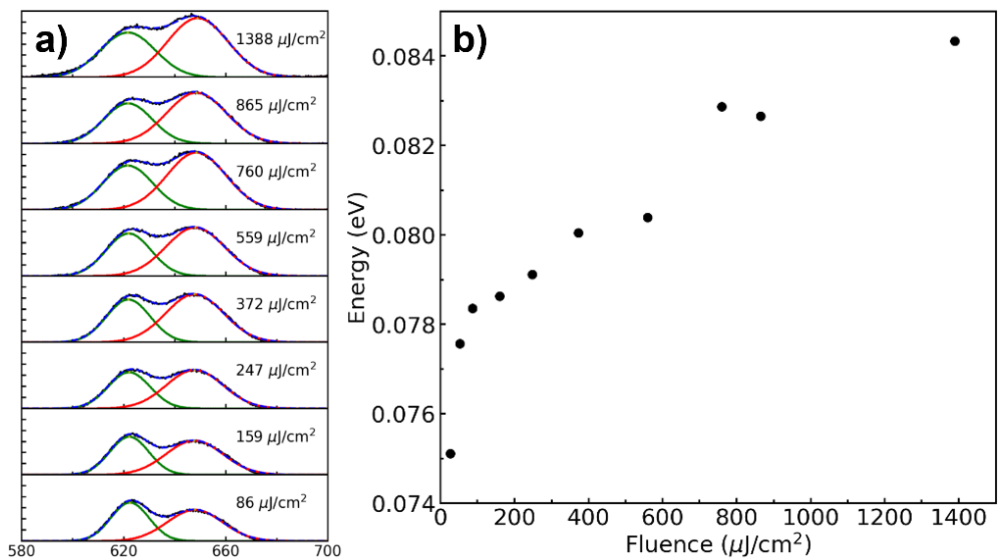

Fig. S8. a) Emission spectra of p-Si $S_8$ array with CdSe/ZnS QDs at different excitation beam fluences, measured at 45° excitation angle and 0° emission angle. b) Energy separation of the peaks of the emission spectra as function of fluence.